\documentclass[prl,floatfix]{revtex4} 

\usepackage{graphicx}
\usepackage{psfrag}
\begin{document}

\def\w{$^{180}W$ }
\def\wn{$^{180}W$}
\def\wn{$^{180}W$n}
\def\a{$\alpha$ }
\def\cw{$CaWO_4$ }
\def\sy{system }
\def\syn{system}
\def\dn{detector}
\def\d{detector }
\def\ds{detectors }
\def\dn{detector}
\def\dsn{detectors}
\def\sc{superconductor }
\def\sa{sapphire }
\def\san{sapphire}
\def\dm{dark matter }
\def\dmn{dark matter}
\def\yy{energy }
\def\yyn{energy}
\def\yys{energies }
\def\yysn{energies}

\def\qf{quenching factor }
\def\qfn{quenching factor}

\title{%
\hbox to\hsize{\normalsize\rm 
\hfil }
\vskip 36pt  Interpretation of Light-Quenching Factor Measurements}
\author{I. Bavykina}
\author{P. Christ}
\author{P. Huff}
\author{J. Ninkovi\'c}
\author{F. Proebst}
\author{W. Seidel}
\author{L. Stodolsky}

\affiliation{Max-Planck-Institut f\"ur Physik 
(Werner-Heisenberg-Institut),
F\"ohringer Ring 6, 80805 M\"unchen, Germany}


\begin{abstract} 

We observe that the pattern of the
\qfn s for  scintillation light 
  from various  ions,  recently studied in $CaWO_4$ in
connection
with \dm \dsn, can be  understood as a saturation phenomenon in
which the
light output  is simply  proportional to track length, independent
of the ion and its \yyn. This observation is in accord with the
high dE/dx
limit of Birks' law. It suggests a simple model for the
intrinsic resolution of light detectors for low energy
 ions, which we briefly discuss.

\end{abstract}

\maketitle

\section{Introduction}

Comparison of  light and heat signals is proving to be an
 effective method of  background reduction in cryogenic \dm
searches
\cite{gs}. In addition,
a detailed understanding of  the reduction of the
light signal   for different recoiling nuclei
may
 be quite  significant for such searches. In a
 \d material containing a variety of nuclei, it 
could permit the  identification of  which type of nucleus has been
struck. This would   allow  study of a possible \dm signal for
different target nuclei in one and the same experimental setup.
Since a simple behavior in the shape of the recoil \yy
spectrum is expected with respect to the mass of the target
nucleus,
this would be a great experimental simplification
and an important check in verifying the \dm signal
\cite{prop}. Furthermore, comparison of the rates on different
nuclei would be useful in learning about the properties of the  \dm
particle\,\cite{dtg}.

\begin{figure}
{{\includegraphics[width=0.5\hsize]
{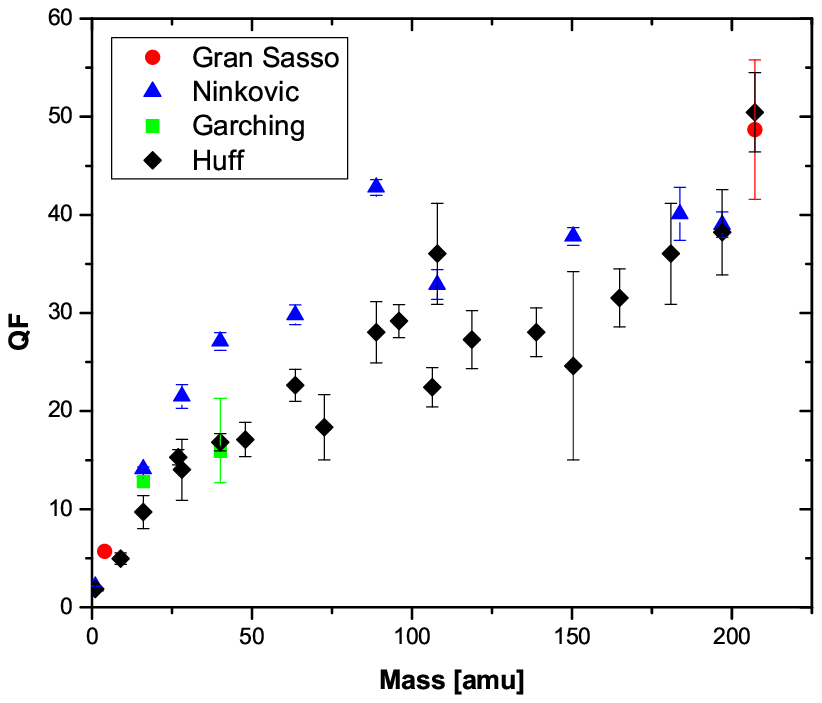}}}
\caption{Measurements of the \qf in  $CaWO_4$.}
\end{figure}
 
\section{ Quenching Factor Curves} 

For these reasons the CRESST collaboration has undertaken a series
of studies by various
methods of the  scintillation  light produced  in $CaWO_4$, the
material presently used in
CRESST. The results are presented in terms of the \qfn, the
reduction factor for the light output for a given ion compared to
that of an electron of the same \yyn. The \qf is a convenient
 quantity  experimentally  since a calibration may be performed by
placing a gamma-ray source, which produces
photoelectrons,  in the same apparatus as one is using  to
study the ions.
This  obviates the need to
understand the absolute efficiencies for light collection and
detection since these will presumably be the same for electron- and
ion- induced scintillation light. On the other hand, the
interpretation then involves
the light output for electrons as well as that for ions.
 
The measurements have been performed at room temperature with ions
projected
onto a  $CaWO_4$ crystal \cite{jel}, \cite{huff}, and with nuclear
recoils induced by neutron scattering \cite{gchg}. In addition  a
low temperature determination was possible using data of a 
prototype
run in 2004 (Run 28) of the CRESST light/heat \ds
 in the Gran Sasso laboratory, through the
identification 
of a group of events as $^{210}$Po decays\,\cite{gs1}. 

 Fig 1 shows these various results.
One observes a rough consistency among the  different methods.
There are two main points to be noted: A) a smooth systematic
increase in the \qf
with the mass or Z of the ion, and  B) an approximate \yy
independence
of the
\qfn, since the various methods involve
ions of different \yysn, ranging from 18\,keV to 2.2\,MeV. The low
temperature (Gran Sasso) results
also suggest
there is not a strong temperature dependence\cite{mikh}.

\begin{figure}
{{\includegraphics[width=0.5\hsize]
{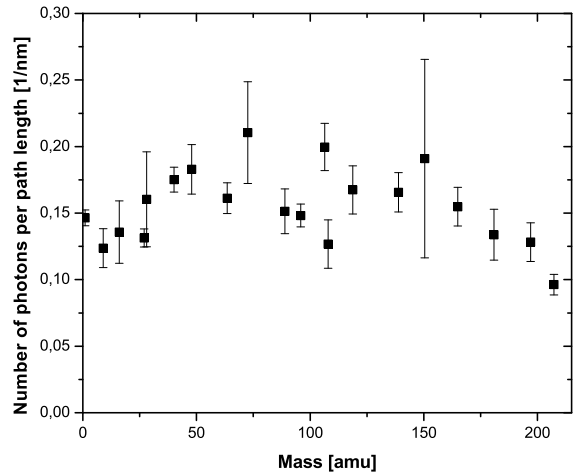}}}
\caption{Photons per unit L for various ions, using
the data of ref\,\cite{huff} at 18 keV and ``path lengths'' L
calculated
from the
SRIM programs as explained in the text. ``Number of photons'' on
the vertical axis simply refers to detected photons, no attempt has
been made to account for efficencies. Error bars are from the
experimental data, fluctuations in the calculation of L are on the
order of a few percent.}
\end{figure}

\section{Proportionality of light to ``track length''}

In this note we would like to point out a simple interpretation
of these data. We start with the question of the dependence on the
type of  ion, that is for data at fixed \yyn.
The \qf  is  the reduction factor with which one divides   the
light output   for an
electron  to find
the light output  for an
 ion of the same \yyn. Thus at fixed \yy the light output for an
ion is  proportional to the
inverse of the \qfn.

\begin{equation}\label{orata}
  light_{ion}={1\over quenching~factor}light_{electron}
\end{equation}

Among the quantities associated with passage of an ion in a
material there are  the ranges or  the track lengths. These
quantities may be found by simulations of the slowing down process,
as from  the SRIM \cite{srim} and STAR \cite{nist} programs. We
shall work with a quantity we call L which we obtain from the SRIM
results  as follows:

\begin{equation}\label{l}
L=\sqrt {R^2_{long} +R^2_{rad}}\;,
\end{equation}
where the R's are the average longitudinal and radial ranges
provided by the simulation. L is intended to provide a
representation of the effective length of the track (see the
remarks in ``Discussion''). Now,  plotting  1/L, one finds a
striking resemblance to  the  pattern
of points in Fig\,1. In view of Eq\,\ref{orata}, this suggests that
the light output for an ion may be
simply proportional to  its ``track length''.

To make this point most effectively we show in Fig\,2 the light
yield  divided by L for different ions, using the
18 keV data of ref\,\cite{huff}. One observes that the points
 scatter around a common value, in agreement with the
idea that
the light output is simply constant per unit  ``length  of track''.
 In this picture of a constant light production per unit length,
one would  assume that there is a saturation value of dE/dx where
this constant value of light output dl/dx is attained,
characteristic of the
material (here \cw)  but independent of the projectile and its
\yyn.

\begin{figure}
{{\includegraphics[width=0.5\hsize]
{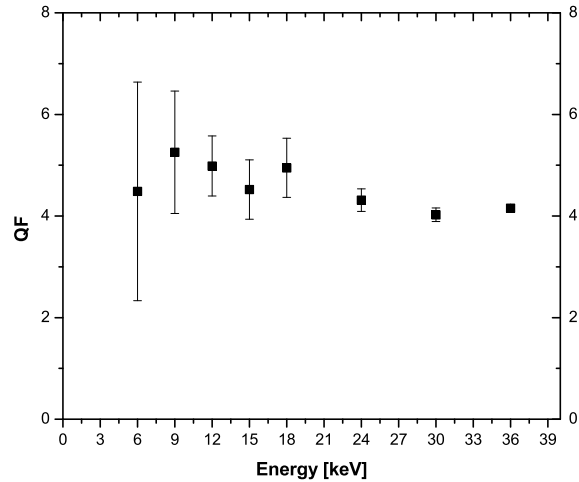}}}
\caption{ Energy dependence of the \qf for Be, from
ref\,\cite{huff}.}
\end{figure}

\section{energy dependence}
As remarked above, a second interesting feature of the
data is the approximate \yy independence of the \qfn.
The data  of Fig\,1 represents a relatively wide span of \yyn,
ranging from 100 keV for the $^{210}$Po
decays  or MeV's for the neutron-induced recoils,  down to 18 keV
for the accelerated ions.      
The fact that these various results are generally in agreement
indicates
that the \qf is roughly \yy independent in this range.
A direct measurement    was also possible with
Be ions using multiply charged states in the ion accelerator
of ref\,\cite{huff}, as shown in Fig 3.
The range of \yys is that relevant for CRESST and most dark matter
searches in general.

If dE/dx is above the saturation value for all
 \yys encountered during the slowing  down process, we naturally
obtain that the light production is proportional to the track
length for any projectile,   and furthermore that it is dependent
on the initial \yy  only through the track length. 
  The track lengths in \cw at these \yys appear in fact to grow
approximately linearly with \yyn, as shown in Fig 4, so that the
light output for ions should also increase approximately linearly
with \yyn.

To finally obtain an \yy independent \qf we require that the light
output for an electron also be linear with the \yyn. 
That this is indeed true  is shown by 
fact that  the electron/photon band in the   light/energy plane, as
reproduced in Fig 5
from ref\,\cite{gs}, is approximately straight. If the band is
straight,
the light yield is proportional to the total \yyn.
 Also, direct tests  with different  gamma-ray
sources have shown the linearity of the scintillation response
\cite{jel}, \cite{huff}, up to 1.3 MeV.

\begin{figure}
{{\includegraphics[width=0.5\hsize]
{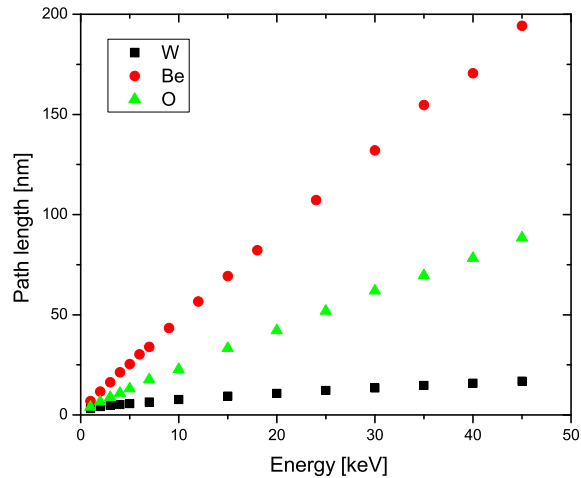}}}
\caption{``Track lengths'' L in \cw calculated from the SRIM
simulation program.}
\end{figure}

\section{Light output from electrons}
It is interesting to inquire where the point for electrons would
lie on Fig 2, that is, how the light production per unit L for
electrons compares to that for ions. Since electrons are
much less ionizing than ions we of course expect the value to be
smaller. We may estimate the ion/electron ratio
 by taking a typical ion, say oxygen,
and using
\begin{equation}\label{orat}
{(light~per~unit~L)_{oxygen}\over
(light~per~unit~L)_{electron}}={light_{oxygen}\over
light_{electron}}{L_{electron}\over L_{oxygen}}={1\over
quenching~factor}{L_{electron}\over L_{oxygen}} \;.
\end{equation}   
 The \qf for  oxygen is 
about 8. At 20 keV   $L_{electron}$ in \cw  is about
$1.7\times 10^3 nm$ \cite{nist}, to be compared with the 40\,nm for
oxygen in Fig 4.
 Thus
\begin{equation}\label{oratb}
{(light~per~unit~L)_{oxygen}\over
(light~per~unit~L)_{electron}}\approx {1\over 8} {1700\over
40}\approx 5 \;.
\end{equation}   
That is to say,   a saturated track in \cw produces about five
times as much light per unit ``length'' as a 20 keV electron does,
when averaged over its whole path;
and so the electron point on Fig 2 would lie a factor five below
the others.
 It appears that while the number of photons
per unit length of track for electrons is of course less than that
for ions, it is not orders of magnitude smaller at these \yysn. 

It is interesting to remark, finally,  that  the approximate
proportionality of light output to \yy  apparently arises for 
electrons and ions  for different reasons. For ions, as explained
above, the saturation of the light output l per unit track length
L, together with the approximate linear growth  of L with \yyn,
gives the proportionality.  For electrons on the other hand  we
have  low dE/dx, where  Birks' law \cite{birks} reads dl$\sim$dE. 
Upon integrating, this gives $l\sim E$, whether the range is
growing linearly with \yy or not.

\begin{figure}
{{\includegraphics[width=0.5\hsize]
{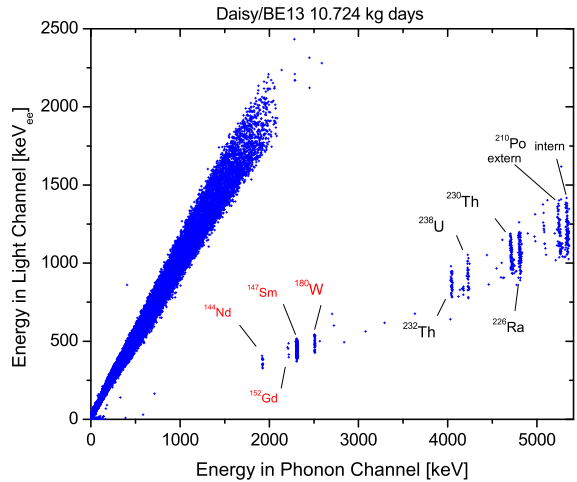}}}
\caption{Events in the \yy (phonon)-light plane, showing alpha
particle and electron-photon bands. From
ref\,\cite{gs}, Fig 7.}
\end{figure}

\section{Discussion}
We have placed the words ``track length'' in  quotation marks
throughout because the exact interpretation of the quantity L,
Eq\,\ref{l}, is not  clear. At these energies it is not evident
that there
is something like a roughly well-defined track for the ion. A low
energy ion in a dense material has a very irregular path and will
create many recoils along the path.  In Fig 6 we show two examples
of tracks, with recoils, from the SRIM program. In the left panel
one sees
that despite the recoils a more or less well-defined track exists.
In the right panel, however the projectile ion has changed
direction radically and the notion of a definite track appears more
doubtful. Furthermore,
the ``track length'' of Eq\,\ref{l} does not account for most
``wiggles'' along the track, and in addition the square root of the
sum of  squares of  averages  is not  the
same as average of a quantity itself. The neglect of ``wiggles''
can have a great effect on estimates of the total track length, as
the  often substantial ``detour factor'' supplied by the STAR
calculations shows.

But on the other hand it is not evident that the track length of
just the incident ion  is the relevant quantity.  A significant
portion of the initial \yy goes into the recoils and 
the saturation region around the track must have some definite
width, so
many of the ``wiggles'' will be ineffective in light production. 
It seems most appropriate to consider L as 
a  measure for the linear dimension of the excited region of the
scintillator; this appear to be an approximate  but nevertheless
useful quantity, as  our results show.

These subtleties should become unimportant at higher \yyn, where we
can comfortably identify L with the ordinary track length. The \yy
where this occurs can be gauged from plots of the ``detour
factor'', which typically approaches one around a few MeV
\cite{nist}.

Turning now to the
 saturation of the  light output, this is 
incorporated in Birks' Law for scintillation  light output per unit
length,
 $dl/dx\sim {dE/dx\over 1+kB dE/dx}$,
in the high $dE/dx$ limit. For inorganic scintillators it is
discussed in Birks \cite{birks}, chapt. 11 and in
Rodnyi \cite{rodnyi}, section 2.3.5. As discussed in these
references  the extensively
studied  NaI-Tl or CsI-Tl scintillators show the transition to the
saturation regime  taking place  for $dE/dx$ between some tens of
MeV/gm\,cm$^2$ and about 100\,MeV/gm\,cm$^2$.

We can try to estimate the saturation dE/dx in our \cw data from
the  fact that at 18\,keV   saturation appears to have occurred for
protons (first point on Fig 2)  but has not for electrons.
 At 18\,keV
 for electrons dE/dx  in \cw is about 8 MeV/gm cm$^2$\cite{nist},
while for 
protons it is 160 MeV/gm cm$^2$ \cite{srim}. We may conclude
 that for \cw then
\begin{equation}\label{dedx}
8 ~MeV/gm \,cm ^2~~~~<~~\biggl({dE\over dx}\biggr)_{saturation}  
  <160~ MeV/gm\, cm ^2 \; .
\end{equation}
 
This span  is rather broad, but of course the saturation turns on
gradually, and the values are in the general
 range  just mentioned.  
 It is interesting to note that at these low \yys the
 dE/dx for electrons  is strongly increasing with decreasing \yyn,
rising to 37 MeV/gm cm$^2$ at  1\,keV. Thus
it is
possible that towards the end of its range the light production
along the electron track  also begins to  saturate.

\begin{figure}
\begin{center}
\begin{tabular}{cc}
{{\includegraphics[width=0.5\hsize]
{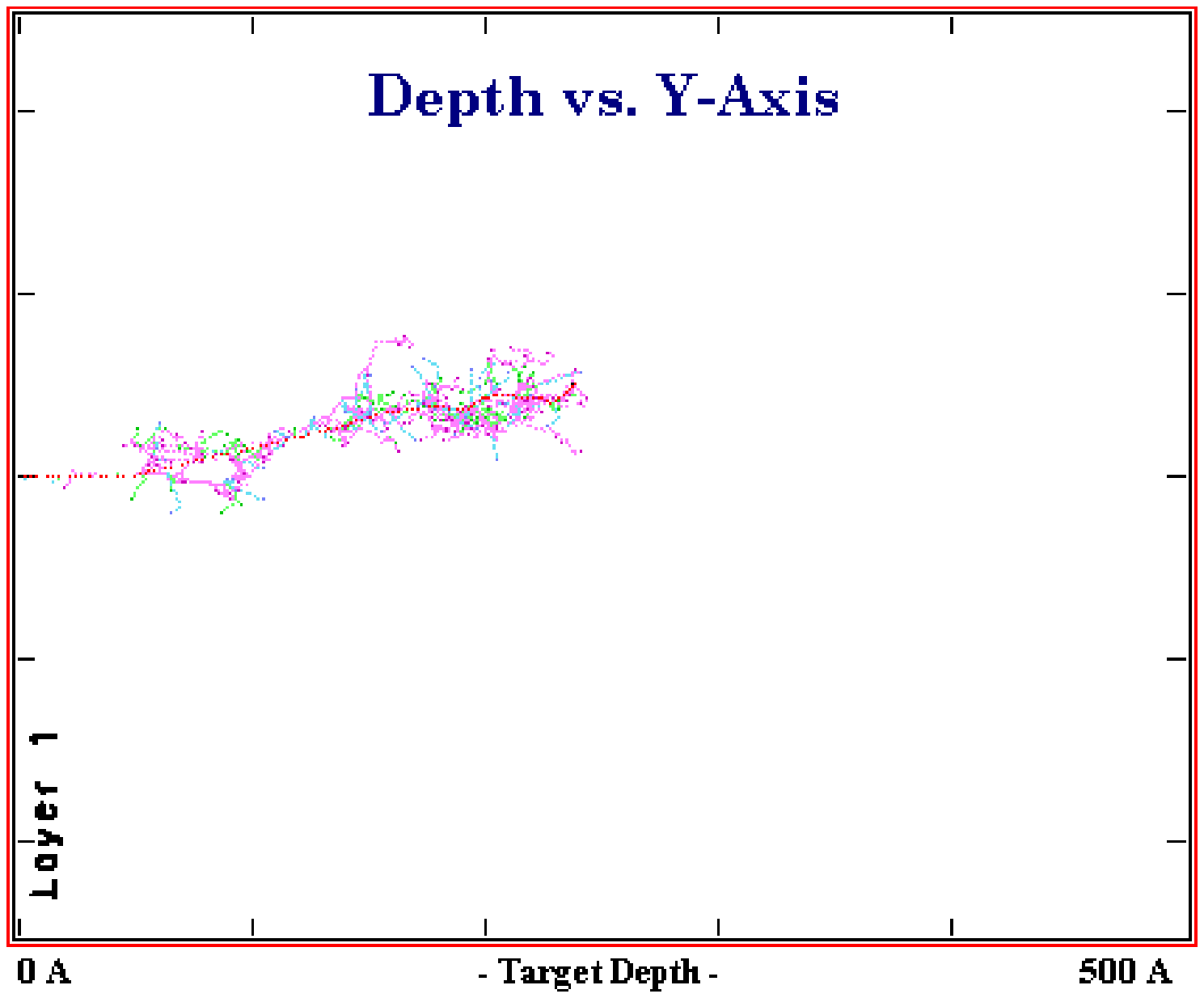}}}&{{\includegraphics[width=0.5\hsize]
{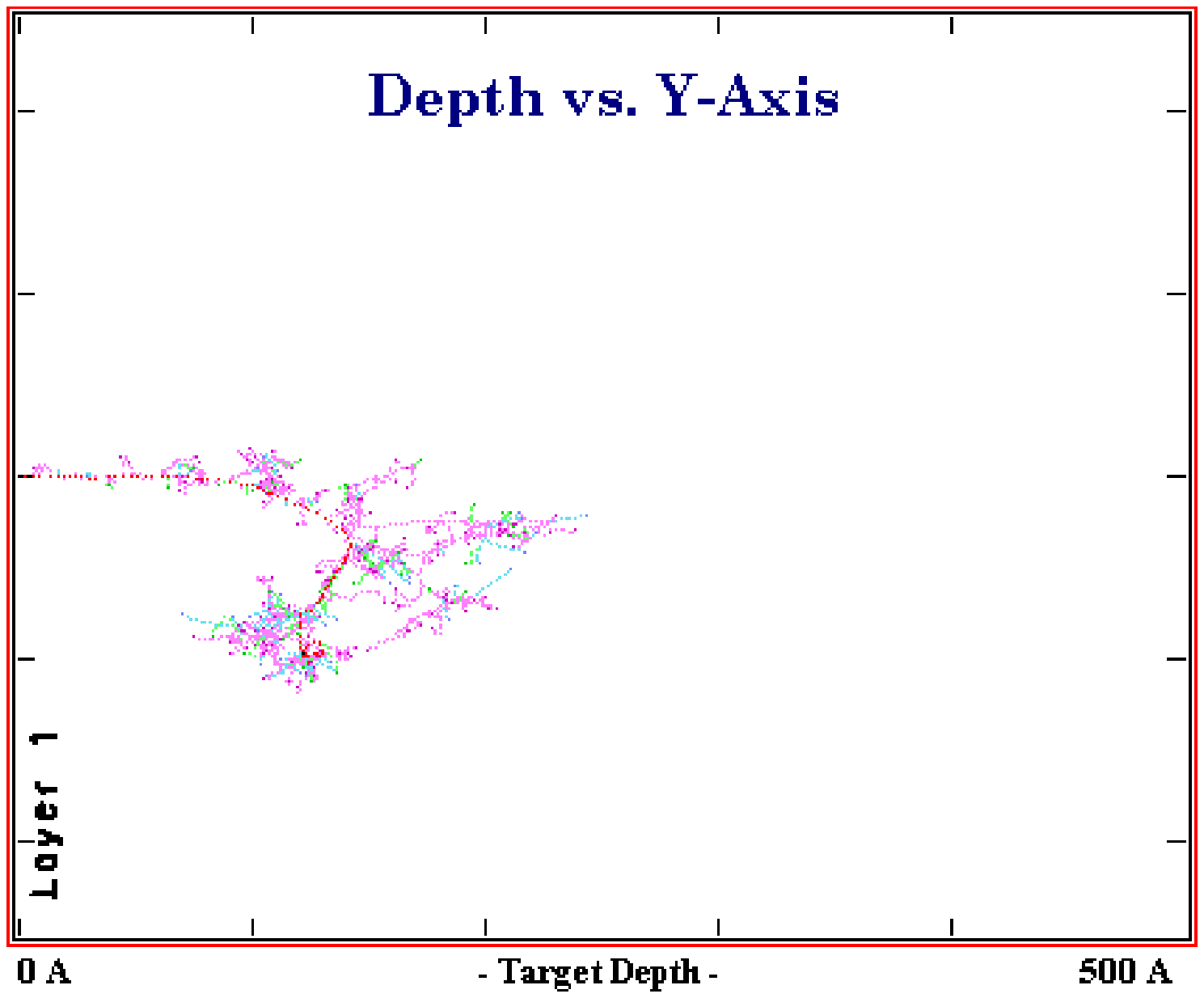}}}
\end{tabular}
\end{center}

\caption{Two examples of tracks of incident ions plus recoil nuclei
produced in $CaWO_4$
by the SRIM simulation. An incoming 20 keV Ca ion is shown in red,
with Ca (green), O(violet) and W (blue) recoils.}
\end{figure}

\section{Comment on Resolution of Light Detectors}

 In any \d system, there will be a limit to the resolution
attainable because there are fluctuations in the underlying
physical process.  In the present case this is the
production of scintillation light.

The picture suggested by our observation that the light output for
ions seems just proportional
to ``track length'' provides  a simple model for these
intrinsic fluctuations:
 the variations in   ``track length'', called  straggling.
On this model the ratio (straggling)/(``track length'') would  
give
the intrinsic fluctuation for the light production by ions.

The SRIM program provides  the fluctuations or  straggling
for the longitudinal and radial  ranges.  For a given ion and \yy
we add these in quadrature, as we did for the range itself  in
Eq\,\ref{l}, and call the result the ``straggle'' \cite{indic}. In
Fig 7 we show the ratio (straggle)/L 
obtained from the program
with Oxygen ions in  $CaWO_4$.   The low \yy value for this
relative fluctuation seems
 quite large. However this is not in disagreement with the data
of Meunier et al, \cite{meunier} where \cw
was irradiated with an Am-Be source,  giving neutron-induced
recoils.
 The  spread of the
nuclear recoil band obtained there also shows 
such large
fluctuations at low \yyn. But evidently   other sources of 
scatter in the data, such as the baseline (electronic)  noise of
the thermometer are present, and it is difficult to say if
intrinsic fluctuations are being observed.
(The nuclear recoil band is presumably dominated by
Oxygen recoils; but simulations for Ca ions would  give  similar
results.)

At  higher \yys  the (straggle)/L ratio found from the simulations 
decreases. This is
a consequence of the fact that while the range continues to
increase
more or
less linearly with \yyn,  the straggling  turns over with
increasing \yyn, growing more slowly than linear, and begins to
resemble
the  $\surd E$ behavior expected from an  accumulation of small
random fluctuations. In the MeV range  (straggle)/L  goes down to
only  some
percent.

For alpha particles at 3 MeV the ratio from the
simulation
is  0.07. This is in marked contrast to the large width of the
alpha
band of Fig 5, which furthermore shows no tendency towards a
$1/\surd E$
behavior.  Presumably the fluctuations are dominated by other
effects such
as inhomogeneities in the  crystal, giving   position-dependent
light production
 and escape. Such a hypothesis is supported by the fact that at the
same average measured light output the electron and alpha bands in
Fig 5 have
closely the same spread. Furthermore we find that
 different crystals have different spreads \cite{widths} in the
light output, while fluctuations of intrinsic origin should be the
same for all  crystals of the same material.

At the present time it thus does not seem  possible to draw any
firm conclusions as to the validity of a straggling model for the
intrinsic resolution. It would be interesting to examine it with
tests using different materials and \yysn. These considerations 
are  of course only relevant when instrumental fluctuations are so
well under control that  intrinsic fluctuations  become the
limiting factor. However they do suggest that at low \yy there may
be a significant intrinsic limit to the attainable resolution. 

\begin{figure}
\includegraphics[width=0.5\hsize]{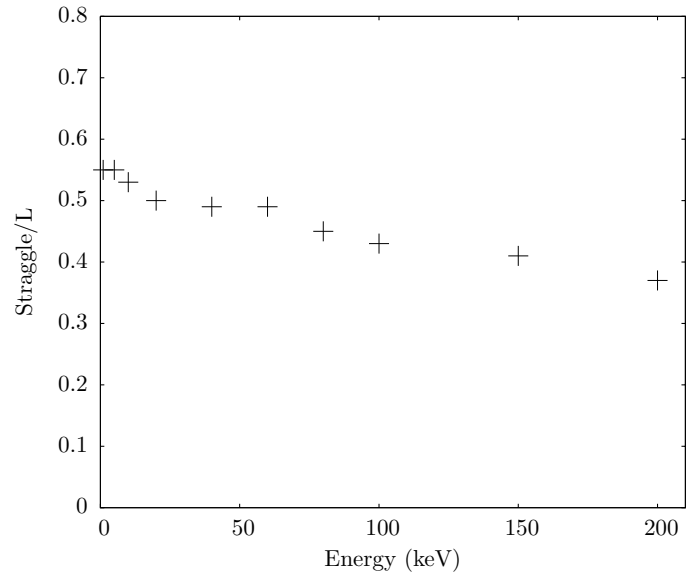}
\caption{ Simulations for the  ratio
(straggle)/L  from  the SRIM program (see text),
for Oxygen ions in \cw . }
\end{figure}

\section{Acknowledgements}
 The calculations were made possible by the availability of the
very useful SRIM programs.


\begin{thebibliography}{00}
\bibitem{indic}  
 In SRIM ``straggle''   is the square root of the
average
square fluctuation. Thus in a Gaussian fit $\sim exp[-
{1\over2}({{\bar l}-l)\over \sigma })^2  ]$  to the event
distribution as a function of $l$ the light output,  
``straggle'' would be equivalent to $\sigma$.
\end{thebibliography}
\enddocument